\documentclass[epj]{svjour}
\usepackage{graphics}
\usepackage{amssymb, amsmath, dsfont, braket}
\usepackage[sort&compress, numbers]{natbib}
\begin{document}
\title{Energy nonconservation as a link between $f(R,T)$ gravity and noncommutative quantum theory}

\author{Ronaldo V. Lobato$^{1,2,3}$ \and
        G. A. Carvalho$^{1,2,3}$
        \and
        A. G. Martins$^{4}$
        \and
        P. H. R. S. Moraes$^{3}$% etc
}                     % Do not remove
\institute{Dipartimento di Fisica and ICRA, 
          Sapienza Universit\`a di Roma, 
          P.le Aldo Moro 5, 
          I--00185 Rome, 
          Italy.
          \and
          ICRANet,
          P.zza della Repubblica 10, 
          I--65122 Pescara, 
          Italy.
          \and
          Departamento de F\'isica,
          Instituto Tecnol\'ogico de Aeron\'autica,
          S\~ao Jos\'e dos Campos, SP, 12228-900,
          Brazil.
          \and
          Departamento de Ci\^encias Naturais,
          Universidade do Estado do Par\'a, 66050-540, Bel\'{e}m, Par\'a,
          Brazil.}
\date{Received: 24-Apr-2018 / Revised version: 02-Aug-2018}
\abstract{
 $f(R,T)$ gravity was proposed as an extension of the $f(R)$ theories, containing not just geometrical correction terms to the General Relativity equations, but also material correction terms, dependent on the trace of the energy-momentum tensor $T$. These material extra terms prevent the energy-momentum tensor of the theory to be conserved, even in a flat background. Energy nonconservation is a prediction of quantum theory with time-space noncommutativity. If time is considered as an operator and there are compact spatial coordinates which do not commute with time, then the time evolution gets quantized and energy conservation can be violated. In the present work we construct a model in a 5-dimensional flat spacetime consisting of 3 commutative spatial dimensions and 1 compact spatial dimension whose coordinate does not commute with time. We show that energy flows from the 3-dimensional commutative slice into the compact extra dimension (and vice-versa), so that conservation of energy is restored. In this model the energy flux is proportional to the energy density of the matter content, leading to a differential equation for $f(R,T)$, thus providing a physical criterion to restrict the functional form of $f(R,T)$. We solve this equation and analyze the behavior of its solution in a spherically symmetric context.
\PACS{
      {04.50.Kd}{Modified theories of gravity}   \and
      {11.10.Nx}{Noncommutative field theory}
     } % end of PACS codes
} %end of abstract
\maketitle
%

%%%%%%%%%%%%%%%%%%%%%%%%%%%%%%%%%%%%%%%%%%%
\section{Introduction}
%%%%%%%%%%%%%%%%%%%%%%%%%%%%%%%%%%%%%%%%%%%
It is known that alternative theories of gravitation offer solutions to some issues that arise when treating General Relativity (GR) as the fundamental theory of gravitation, such as the late-time acceleration of the Universe expansion \cite{Neveu2017, Barrow2007} and flat rotation curve of galaxies \cite{Albada1986, Swaters2000}. Those aforementioned problems are solved in GR by assuming the presence of some kind of energy and matter whose properties are unusual and unknown, therefore being named dark energy and dark matter, respectively. Such parametrization is the so-called $\Lambda$CDM (Lambda cold dark matter) model.

Despite dark energy is well described in GR by a cosmological constant $\Lambda$, some issues related to the $\Lambda$CDM model are frequently brought to the literature, such as the cosmological fine-tuning problem and some small scale problems \cite{DelPopolo2017}. In this context, scalar field theories of gravity in the framework of string theory and supergravity were proposed as candidates to describe the cosmic acceleration. However, observational data do not favor these scalar theories until now \cite{Sotiriou2010}.

Even though GR turned out to be quite successful in many levels, it remains necessary a theoretical framework in which the early-time and late-time accelerations are described in a consistent way. There are many works in this direction, using different theoretical approaches. For a complete review, see \cite{nojiri2017}. Besides, even if the current modified gravity proposals  do not lead to the ultimate  theory of gravity, there is a plethora of phenomena whose explanation goes beyond General Relativity and the Standard Model of particle physics. Therefore, this is a rich field to explore if we want to achieve a description of the Universe consistent at both astrophysical and cosmological levels \cite{nojiri2017}.

An approach to alternative gravity consists in modifying the Einstein-Hilbert action, replacing the Ricci scalar $R$ by some function $f(R)$. In fact, $f(R)$ gravity is one of the most known alternative theories of gravity and although its geometric corrections to GR can solve the cosmic acceleration and galactic astronomy issues \cite{Capozziello2008a}, with no need for dark energy and dark matter, some shortcomings of the theory arise from Solar System tests, as indicated in \cite{nojiri2003, nojiri2007a}. 

$f(R)$ gravity is a family of theories, each one being defined by a specific function $f$. In some cases the form of such function may be inspired by corrections coming from quantum gravity or string/M-theory (see \cite{sotiriou/2009} and references therein). $f(R)$ gravity can also be generalized by means of non-minimal matter-geometry coupling. As shown in \cite{nojiri/2007b}, this modified theory can be used to describe the occurrence of a transient phantom era, thus taking into account the late-time acceleration. It was also shown that some non-linear couplings may help to solve the coincidence problem \cite{nojiri/2007b}. Therefore, a wide range of phenomena can be explained by $f(R)$ gravity and its generalizations, and some of them successfully pass Solar System tests.

Another possibility is to consider the action as a function of both the Ricci scalar and the trace of the energy-momentum tensor $T$, the $f(R,T)$ gravity, proposed by Harko {\it et al} \cite{harko2011}. The $T$-dependence of this theory was originally inspired by the introduction of quantum effects and by supposing the existence of exotic imperfect fluids. Both assumptions can indeed account for a nonconservative energy-momentum tensor, as predicted by the $f(R,T)$ theory of gravity. We also mention that this theory has been generalized to the so-called $f(R,T, R_{\mu\nu}T^{\mu\nu})$ \cite{odintsov2013}. The $f(R,T)$ gravity has passed Solar System tests, as explained in \cite{deng2015, shabani2014}, and was successfully applied to the study of compact astrophysical objects \cite{Moraes2016, Carvalho2017d, Sharif2018}. It also predicts a matter-geometry coupling that yields geodesic motion deviations \cite{harko2011}. As we will discuss later, the non-nullity of the covariant derivative of the energy-momentum tensor in $f(R,T)$ gravity leads to a violation of energy and momentum conservation laws, which may be related to quantum phenomena, such as a possible time-space noncommutativity.

In quantum physics the notion of noncommutativity plays a crucial role. The canonical quantization of a classical theory consists in replacing the position coordinates and their conjugate momenta by Hermitian operators $\hat x_i$ and $\hat p_j$ obeying the Heisenberg commutation relation, $[\hat{x}_i, \hat{p}_j] = i\hbar\delta_{ij}$. This relation induces uncertainties in simultaneous measurements of position and momentum, so that the phase space becomes discretized with a minimum area of $\hbar/2$. According to \cite{Jackiw2002}, in 1930 Heisenberg raised the notion of noncommuting spatial coordinates, suggesting that a coordinate uncertainty principle could help to deal with the problem of self-energies, i.e., the ultraviolet divergences occurring in quantum field theory. Even though the suggestion was explored in some works (see \cite{Snyder1947}, for example), eventually the idea was abandoned due to the success of the renormalization procedure in quantum electrodynamics \cite{Szabo2003}. Nowadays it is known that the original motivation have proved to be wrong: noncommutative quantum theories are not divergence free and need to be renormalized \cite{Chaichian2000}. 

The notion of noncommutative spacetime was revived in the context of quantum field theory \cite{Doplicher1995}. High precision simultaneous measurements of spatial coordinates would require high energies, which in turn could produce a horizon, according to the GR predictions. This should break down the picture of spacetime as a manifold at distances of the order of Planck length, $l_{\rm P} = \sqrt{G\hbar/c^3}\approx 10^{-33}$ cm \cite{Doplicher1995, Snyder1947}, the scale which presumably marks the transition from classical to quantum regime of spacetime. 

Modifying the spacetime at short distances might be relevant for quantum gravity theories too, and the subject was indeed considered in the context of string theory, where noncommutative Yang-Mills theory arises as a limit \cite{Connes1998, Seiberg1999, Szabo2003}. 

The connection between noncommutative physics and gravity have already been considered in the literature, and we can mention at least two approaches to it. The first one consists in considering corrections to the metric and the other one corresponds to modifying the energy-momentum tensor \cite{NICOLINI2009a}, where the noncommutative effects come out as a smeared density $\rho_{\theta}=M/(4\pi\theta)^{3/2}\exp(-r^2/4\theta)$ \cite{Nicolini2006}, with $\theta$ being the noncommutative parameter and $M$ the mass. This distribution was used in some works concerning black holes \cite{Mehdipour2012}, wormholes \cite{Sharif2013, Rahaman2015a}, gravastars \cite{Banerjee2016, Ovgun2017} and alternatives theories of gravity \cite{Jamil2014, Sharif2015, Rani2016, Sharif2017a}.

In the present manuscript we propose a connection between the nonconservation of the energy-momentum tensor in $f(R,T)$ gravity and the violation of energy conservation in scattering processes, an effect predicted by the noncommutative quantum theory with time operator and compact noncommutative dimensions. Scattering processes are studied in particle physics, where there has been a search for physics beyond the standard model, in particular the existence of extra dimensions. One way to do this search is through particle collisions in experiments like those performed at the Large Hadron Collider (LHC). Indeed, the possibility of a violation of energy-momentum conservation has been considered in \cite{ducloue2014}. There have been many attempts to estimate these effects \cite{delduca1995, orr1997, kwiecinski1999, khachatryan2015, datta2014}, and even though there is no direct evidence of this phenomenon  up to now, it has not been ruled out for all energies achieved by LHC \cite{khachatryan2015, datta2014}, leaving the energy-momentum nonconservation as an open question.

The article is organized as follows. In Sec. \ref{sec:frt}, we show the non-nullity of the covariant derivative of the energy-momentum tensor in $f(R,T)$ gravity and obtain a modified continuity equation, which shows that energy is not conserved in the ordinary 4-dimensional spacetime. In Section \ref{sec:ncqt}, we recall some basic aspects of noncommutative quantum mechanics in 2 dimensions (noncommutative plane and noncommutative cylinder) and extend the main results to a 5-dimensional spacetime which contains the noncommutative cylinder as a subspace. In section \ref{sec:extra}, we show that these two instances of energy nonconservation can be related to each other, leading to a restoration of the principle of conservation of energy in the 5-dimensional scenario. In Sec. \ref{sec:conc} we present our conclusions.

%%%%%%%%%%%%%%%%%%%%%%%%%%%%%%%%%%%%%%%%%%%
\section{The $f(R,T)$ gravity}\label{sec:frt}
%%%%%%%%%%%%%%%%%%%%%%%%%%%%%%%%%%%%%%%%%%%

The $f(R,T)$ gravity \cite{harko2011} is a generalization of the $f(R)$ theories (check, for instance, \cite{Nojiri2011}). Its starting point is a gravitational action that depends on an arbitrary function of both the Ricci scalar $R$ and the trace of the energy-momentum tensor $T$. The dependence on $T$ was inspired initially by the consideration of existence of exotic imperfect fluids and quantum effects, which are sometimes associated with conformal anomaly \cite{harko2014, Harko2015a, Liu2016, Zaregonbadi2016a}.

From now on, we consider the metric signature to be $+,-,-,\cdots,$ and adopt Planck units ($\hbar=c=G=1$). 
The $f(R,T)$ action thus reads

\begin{equation}\label{acao}
  \mathcal{S}=\int d^{4}x\sqrt{-g}\left[  \frac{f(R,T)}{16\pi}+\mathcal{L}_{m}\right],
\end{equation}
where $f(R,T)$ is some function of $R$ and $T$, $\mathcal{L}_{m}$ is the matter Lagrangian density and $g$ is the determinant of the metric tensor $g_{\mu\nu}$. 

The field equations of $f(R,T)$ gravity present extra terms which force the covariant derivative of the energy-momentum tensor to be non-vanishing:

\begin{equation}
  \label{divtensor}
  \nabla^{\mu}T_{\mu\nu}=\frac{f_{T}(R,T)}{8\pi+f_{T}(R,T)} \left[(\mathcal{L}_mg_{\mu\nu}-T_{\mu\nu})\nabla^{\mu}\ln f_{T}(R,T)+\nabla^{\mu} \left(\mathcal{L}_m-\frac{1}{2}T\right)g_{\mu\nu}\right],
\end{equation}
where it was used the notation $f_T(R,T)=\partial f(R,T)/\partial T$. This intriguing feature of this theory, as well as the dependence on $T$, was studied considering particle creation in a 4-dimensional Universe \cite{harko2014}. In the present paper we show that the non-nullity of the 4-divergence of $T^{\mu\nu}$ can be seen as a consequence of the existence of a compact spatial extra dimension. This is interesting, since extra dimensions have been explored in a variety of theoretical studies, and its existence remains as an open question from the experimental point of view. 
In this context, the 4D $f(R,T)$ gravity can be seen as an effective theory, which arises when the compact extra dimension is considered to be very small. Just as an example, one could consider the following five-dimensional metric tensor, spherically symmetric in the three-dimensional ordinary spatial slice \cite{emparan/2008},
\begin{equation}
ds^2 = e^{a(r)}dt^2 - e^{b(r)}dr^2 - r^2d\Omega^2 - \mathcal{R}^2(r) dy^2,
\end{equation}
where $y$ is an angular coordinate used to parametrize the compact dimension and $\mathcal{R}(r)$ is the corresponding compactification radius. The extra dimension length scale can be treated as a local quantity, and an observer is not aware of its existence. In the limit of $\mathcal{R}(r)\rightarrow \mathcal{R_{\rm cte}}$ and $r\rightarrow\infty$ we have a product of a flat Lorentz geometry and a circle.

The effective (4D) metric  $g_{\mu \nu }$ is the well-known Schwarzschild geometry (interval $g_{\mu \nu }dx^{\mu }dx^{\nu }$) and the four-dimensional energy density and pressure can be described by an effective $T^{\mu\nu}$. Supposing that the energy-momentum tensor describes a perfect fluid without pressure, i.e., dust, we can assume that the matter Lagrangian density is  given by $\mathcal{L}_m=\rho$. 

In General Relativity the energy-momentum tensor is covariantly conserved, so that a flat spacetime background yields 
well-defined conservation laws, in particular described by the continuity and Navier-Stokes equations. Therefore, it is natural to consider the Minkowski metric as the background spacetime metric, so that $\nabla_\mu T^{\mu\nu}=\partial_\mu T^{\mu\nu}$. For $\nu=0$ equation \eqref{divtensor} leads to the following modified continuity equation:
\begin{equation}\label{noncont}
\frac{\partial \rho}{\partial t}+\nabla \cdot\vec{j}=\frac{f_T}{8\pi +f_T}\left[\frac{1}{2}\frac{\partial \rho}{\partial t}-\vec{j}\cdot\vec{\nabla}(\ln f_T)\right],
\end{equation}
where $\rho$ is the energy density of the dust and $\vec{j}$ represents its energy current density. 
We remark that the weak field limit was successfully studied in some earlier papers, such as, for example, \cite{nojiri/2004, harko2010, nojiri2010, harko2011, odintsov2013, harko2014a, deng2015, haghani2013}. In this way, our approach gets justified.

Equation \eqref{noncont} shows that in $f(R,T)$ gravity the energy conservation law is not respected. This can be interpreted as an energy flux that is ejected (injected) from (in) the system. The loss or gain of energy will depend on the function $f_T(R,T)$. From \eqref{noncont}, for $f_T=0$, the standard continuity equation is recovered.

%%%%%%%%%%%%%%%%%%%%%%%%%%%%%%%%%%%%%%%%%%%
\section{Noncommutative quantum mechanics}\label{sec:ncqt}
%%%%%%%%%%%%%%%%%%%%%%%%%%%%%%%%%%%%%%%%%%%

Canonical non-commutative $D$-dimensional space-time can be constructed by replacing the coordinates $x_{i}$ by Hermitian operators $\hat{x}_{i}$, which obey the Heisenberg-like commutation relations 
$[\hat{x}_i,\hat{x}_j] = i\theta_{ij}\mathds{1},
$, where $\theta$ is an antisymmetric constant tensor and $\mathds{1}$ is the unity element \cite{Szabo2003, Douglas2001}. The noncommutative spacetime corresponds to the noncommutative algebra $\mathcal{A}_{\theta}\left(\mathbb{R}^D\right)$ generated by the set $\{\hat{x}_1\,,\,\dots\,,\,\hat{x}_D\,,\,\mathds{1}\}$. In such a framework the notion of point is no longer meaningful, since pairs of non-commuting coordinates cannot be simultaneously measured. 
The parameters $\theta_{ij}$ determine the size of the fundamental discretization cells, and the uncertainties $\Delta x_i$ are related to each other through inequalities which resemble the usual uncertainty relations of ordinary quantum mechanics \cite{Doplicher1995}.

Some of the most interesting physical consequences of spatial noncommutativity are the nonlocal effects and the breakdown of Lorentz invariance \cite{Seiberg2000}. However, one can also postulate a noncommutativity between time and space ($\theta_{0i}\neq 0$ in commutation relations). Even though there have been claims that the $S$-matrix of such theories would not be unitary \cite{Chaichian2001}, in \cite{Doplicher1995} it was demonstrated how to construct unitary quantum field theories on canonical noncommutative spaces. Following the same lines, \cite{Balachandran2004a} developed unitary quantum mechanics in the noncommutative plane. 
In \cite{Balachandran2004} and \cite{Balachandran2007} the consequences of a compact spatial coordinate which does not commute with time have been studied. As shown in \cite{Balachandran2007}, this hypothesis leads to a remarkable effect: the nonconservation of energy. 

In this work we consider a 5-dimensional spacetime $\mathcal{M}_{\theta}$, whose spatial section is topologically equivalent to $\mathbb{R}^3\times S^1$, where $S^1$ denotes a circle, and such that the time operator ($\hat{x}_0$) do not commute with the compact coordinate ($\hat{x}_4$). These two noncommuting coordinates generate the noncommutative cylinder \cite{Chaichian2001}, being responsible for the discretization of time. The commutative 3-dimensional subspace has coordinates $\hat x_1$, $\hat x_2$ and $\hat x_3$, and the corresponding 3-dimensional interactions remain unchanged. These commutative coordinates generate the algebra of functions on $\mathbb{R}^3$, denoted by $\mathcal{F}\left(\mathbb{R}^3\right)$.  

%%%%%%%%%%%%%%%%%%%%%%%%%%%%%%%%%%%%%%%%%%%
\subsection{Noncommutative plane}
%%%%%%%%%%%%%%%%%%%%%%%%%%%%%%%%%%%%%%%%%%%

As a preparation to study the noncommutative cylinder we first review the construction of the noncommutative plane, which is the simplest canonical noncommutative space, also known as Moyal plane \cite{Balachandran2004}. It corresponds to the algebra $\mathcal{A}_{\theta}\left(\mathbb{R}^2\right)$ generated by $\hat{x}_0$, $\hat{x}_4$ and $\mathds{1}$, obeying the commutation relation

\begin{equation}\label{ncomuta}
[\hat{x}_0, \hat{x}_4]=i\theta\mathds{1}.
\end{equation}
where $\theta$ represents the noncommutative parameter.

Since equation \eqref{ncomuta} is formally equivalent to the Heisenberg algebra of a single particle in a line, it can be realized as operators acting on the Hilbert space $L^2\left(\mathbb{R}^2\right)$, the so-called Schr\"{o}dinger representation. The uncertainties $\Delta x_0$ and $\Delta x_4$ satisfy a Heisenberg-like inequality, that is,
\begin{equation}
\Delta{x_0}\Delta{x_4}\geq{\frac{\theta}{2}}.
\end{equation}

In order to incorporate an adequate notion of time displacement, it is necessary to introduce an analogue of the ordinary time derivative acting on the time operator. Let us denote by $\hat{P}_0$ the operator defined by $\hat{P}_0\hat{\alpha}=-\frac{1}{\theta}[\hat{x}_4,\hat{\alpha}]$, which plays the role of $i\partial_0$ in this algebraic context.
Time displacement (by an amount of $t$) is generated by the operator $e^{-it\hat{P}_0}$. It follows that
\begin{equation}\label{ncp}
e^{-it\hat{P}_0}\hat{x}_0=\hat{x}_0+t.
\end{equation}

The quantum mechanics in the Moyal plane is the theory of Hermitian operators acting on the Hilbert space of physical states, which are the solutions of the algebraic Schr\"{o}dinger equation \cite{Balachandran2004a}:
\begin{equation}\label{Psi}
\hat{H}(\hat{x}_4)\hat{\psi}(\hat{x}_4,\hat{x}_0)=\hat{P}_0\hat{\psi}(\hat{x}_4,\hat{x}_0).
\end{equation}

%%%%%%%%%%%%%%%%%%%%%%%%%%%%%%%%%%%%%%%%%%%
\subsection{Noncommutative cylinder} 
%%%%%%%%%%%%%%%%%%%%%%%%%%%%%%%%%%%%%%%%%%%

Since the topology of the extra dimension is a circle, the functions $\hat\psi(\hat x_0,\hat x_4)$ must be periodic in $\hat x_4$ with period equal to $t_{\mathcal{R}}=2\pi \mathcal{R}$, where $\mathcal{R}$ is the circle radius. Therefore, only integral powers of $e^{i\frac{\hat x_4}{\mathcal{R}}}$ are allowed, and the usual Fourier techniques can be used to construct elements of the algebra $\mathcal{A}_{\theta}\left(\mathbb{R}\times S^1\right)$ of the noncommutative cylinder. The appropriate commutation relation (which can be obtained from equation \eqref{ncp}) reads

\begin{equation}\label{nccy}
\left[\hat{x}_0,e^{\frac{i\hat{x}_4}{\mathcal{R}}}\right]=-\frac{\theta}
{\mathcal{R}}e^{i\frac{\hat{x}_4}{\mathcal{R}}}.
\end{equation}

By using the Baker-Campbell-Hausdorff formula \cite{Hall2000} one can show that

\begin{equation}
e^{i\frac{\hat x_4}{\mathcal{R}}}e^{i\frac{2\pi \mathcal{R}}{\theta}\hat x_0}=e^{i\frac{2\pi \mathcal{R}}{\theta}\hat x_0}e^{i\frac{\hat x_4}{\mathcal{R}}}.
\end{equation}
Therefore, the operator $e^{i\frac{2\pi \mathcal{R}}{\theta}\hat x_0}$ commutes with any element of the noncommutative cylinder. By the Schur's lemma \cite{Hall2000}, it is proportional to the identity operator $\mathds{1}$, that is,
\begin{equation}\label{phase}
e^{i\frac{2\pi \mathcal{R}}{\theta}\hat x_0}=e^{i\varphi}\mathds{1}.
\end{equation}
The phase constant $e^{i\varphi}$ classifies the irreducible representations of $\mathcal{A}_{\theta}\left(\mathbb{R}\times S^1\right)$, as explained in \cite{Balachandran2004}. If $|x_0\rangle$ denotes the eigenvector of $\hat x_0$ with eigenvalue $x_0$, then it follows from equation (\ref{phase}) that

\begin{equation}
e^{i\frac{2\pi \mathcal{R}}{\theta}\hat x_0}\ket{x_0} = e^{i\frac{2\pi \mathcal{R}}{\theta}x_0}\ket{x_0}=e^{i\varphi}\ket{x_0},
\end{equation}
leading to the conclusion that the spectrum of $\hat x_0$ gets quantized, with step $\tau_{0}=\frac{\theta}{\mathcal{R}}$, that is,

\begin{equation}\label{discrete_time}
\text{spec}\hat x_0=\left\{\frac{\theta}{\mathcal{R}}\left(m+\frac{\varphi}{2\pi}\right)\quad,\quad m\in\mathbb{Z}\right\}.
\end{equation}

%%%%%%%%%%%%%%%%%%%%%%%%%%%%%%%%%%%%%%%%%%%
\section{Energy nonconservation in noncommutative quantum theory and $f(R,T)$ gravity}\label{sec:extra}
%%%%%%%%%%%%%%%%%%%%%%%%%%%%%%%%%%%%%%%%%%%

According to the limitations imposed by equation (\ref{discrete_time}), time translations on the noncommutative cylinder are restricted to discrete jumps generated by $e^{-in\tau_0\hat{P}_0}$. For a time-independent Hamiltonian the equation (\ref{Psi}) must be replaced by (see \cite{Balachandran2004,Balachandran2007})

\begin{equation}\label{discrete_evolution}
e^{-in\tau_0\hat{P}_0}\hat{\Psi}(\hat{x}_4,\hat{x}_0)=\hat{\Psi}(\hat{x}_4,\hat{x}_0+n\tau_0)=e^{-i n\tau_0\hat{H}(\hat{x}_4)}\hat{\Psi}(\hat{x}_4,\hat{x}_0).
\end{equation}

Equation (\ref{discrete_evolution}) implies that the time-evolution in $\mathcal{A}_{\theta}\left(\mathbb{R}\times S^1\right)$ is discrete, being given by the action of the unitary operator $\hat{U}(n\tau_0)=e^{-in\tau_0\hat{H}}$. Therefore, the generator of time-evolution (the Hamiltonian) is defined only up to discrete shifts by multiples of the quantum $\frac{2\pi}{\tau_0}$. The physical consequences of this result were studied in \cite{Balachandran2007}, where it was shown that conservation of energy would be violated in scattering processes. 

Let us now consider a system of particles moving in the 5-dimensional spacetime $\mathcal{M}_{\theta}$ described earlier. At the algebraic level we have $\mathcal{M}_{\theta}=\mathcal{F}\left(\mathbb{R}^3\right)\otimes\mathcal{A}_{\theta}\left(\mathbb{R}\times S^1\right)$, where the symbol $\otimes$ denotes the tensor product of algebras. Each particle is under the action of some time-independent Hamiltonian of the form $\hat{H}_{\text{full}}=\hat{H}_{\text{3D}}(\hat{x}_1,\hat{x}_2,\hat{x}_3)+\hat{H}_{\text{extra}}(\hat{x}_4)$, so that the time-evolution operator takes the form $\hat{U}_{\text{full}}=\hat{U}_{\text{3D}}\hat{U}_{\text{extra}}$. Therefore, the above considerations concerning the time-evolution in the noncommutative cylinder remains valid for $\mathcal{M}_{\theta}$, that is, the time-space noncommutativity in $\mathcal{A}_{\theta}\left(\mathbb{R}\times S^1\right)$ induces a time discretization in $\mathcal{M}_{\theta}$. On the other hand, since the Hamiltonian does not contain any coupling between the extra dimension and the ordinary ones, the 3-dimensional interactions are left unchanged. 

For each particle of the system the transition probability from an initial state $\ket{E_{i}}$ to a final state $\ket{E_{f}}$ will be denoted by $T_{i\rightarrow f}$. The allowed transitions are those for which $E_{f}-E_{i}=\frac{2\pi m}{\tau_0}$, with $m\in \mathbb{Z}$, and according to \cite{Balachandran2007} the transition rate $T_{i\rightarrow f}$ can be factorized as follows:

\begin{equation}\label{delta}
T_{i\rightarrow f}(E_i\,,\,E_f) = \Gamma_{i\rightarrow f}(E_i\,,\,E_f) \delta_{S^1}(E_f-E_i),
\end{equation}
where $\delta_{S^1}(E_f-E_i)$ denotes the periodic Dirac delta distribution with period $\frac{2\pi}{\tau_{0}}$, meaning that energy is not a conserved quantity. In general the factor $\Gamma_{i\rightarrow f}(E_i\,,\,E_f) $ is a decreasing function of $|E_f-E_i|$. 

The rate of energy transfer $\Phi_{i \rightarrow f}$ associated to the transitions from states with energy $E_{i}$ to states with energy $E_{f}$ is proportional to the energy density $\rho_i(E_i)$. Besides, it follows from \eqref{delta} that 

\begin{equation}\label{fluxo_i}
\Phi_{i \rightarrow f}=(E_f-E_i)\Gamma_{i\rightarrow f}(E_i\,,\,E_f) \delta_{S^1}(E_f-E_i)\rho_i(E_i).
\end{equation}
Taking into account all possible final energies for these transitions and making use of \eqref{fluxo_i} we can write this partial energy flux as

\begin{equation}%\nonumber
\Phi_{i}=\frac{2\pi }{\tau_0}\sum_{m=-\infty}^{\infty}m\Gamma_{i \rightarrow f}\left(E_i\,,\,E_i+\frac{2\pi m}{\tau_0}\right)\rho_i(E_i).
\end{equation}

At this point we propose a physical interpretation for the nonconservation of energy caused by the quantum transitions described by equation \eqref{delta}: we suppose that the energy flows from the extra dimension into the ordinary 3-dimensional spatial slice, and vice-versa. Energy is thus conserved in the full (5-dimensional) space-time. Consequently, it would be interesting to estimate the total rate of energy transfer, taking into account all possible transitions of a simple system of particles. 
       
Let $N$ denote the number of particles of the system, supposed to be large enough to make classical effects appear. This system can be seen as a ``fluid'' consisting of many particles which are constantly changing their energy levels. Thus, the transitions occurring at the quantum level induce a continuous variation of the mean energy density, from the macroscopic point of view.

In order to take into account the contributions of all possible initial energies we suppose that the (statistical) probability to find a particle with initial energy $E_i$ is given by some equilibrium distribution $P_{i}(E_i)$ and sum over all the possible initial energies using this weight, that is, 

\begin{eqnarray}
\Phi &=& N\sum_i\Phi_i P_i(E_i)\nonumber\\
\Phi &=& \frac{2\pi N }{\tau_0}\sum_{i}\sum_{m=-\infty}^{\infty}m\Gamma_{i \rightarrow f}\left(E_i\,,E_i+\frac{2\pi m}{\tau_0}\right)\rho_i(E_i)P_i(E_i).
\end{eqnarray}

Considering the special case of an interaction whose transition matrix is approximately independent of the initial energy we can extract $\Gamma_{i \rightarrow f}$ from the sum over $i$, thus leading to

\begin{equation}\label{energy_density}
\Phi \approx \frac{2\pi N }{\tau_0}\sum_{m=-\infty}^{\infty}m\Gamma_{i \rightarrow f}(m)\sum_i\rho_i(E_i)P_i(E_i) = \frac{2\pi N \xi}{\tau_0}\overline{\rho},
\end{equation}

where $\overline{\rho}$ is the average energy density and the sum over $m$ was denoted by $\xi$. This result shows that the energy flux $\Phi$ depends on the ratio $\tau_0=\frac{\theta}{\mathcal{R}}$, so that even if $\theta\longrightarrow 0$ and $\mathcal{R} \longrightarrow 0$, we still can get nontrivial results (finite $\tau_0$).

We argue that in the limit of large values of $N$ the average energy density appearing in equation \eqref{energy_density} can be identified with the energy density of the fluid, that is, $\Phi=k(N,\tau_0)\rho$. 

According to our interpretation, the non-nullity of the right-hand side of equation (\ref{noncont}) is a consequence of the energy flux from the 3-dimensional commutative subspace to the extra dimension (and vice-versa). Therefore, we can write

\begin{equation}\label{noncont+rho}
\frac{\partial \rho}{\partial t}+\nabla\cdot \vec{j}=k\rho
\end{equation}
and
\begin{equation}
\label{diff_eq_f}
\frac{f_T}{8\pi +f_T}\left[\frac{1}{2}\frac{\partial \rho}{\partial t}-\vec{j}\cdot\vec{\nabla}(\ln f_T)\right] = k\rho.
\end{equation}

Equation \eqref{diff_eq_f} corresponds to the following differential equation for $f_T$:
\begin{equation}\label{diff}
\vec{j}\cdot\vec\nabla f_T + \left(k\rho-\frac{1}{2}\frac{\partial \rho}{\partial t}\right) f_T = -8\pi k\rho.
\end{equation}

Now, based on physical grounds, we assume that the energy density is spherically symmetric, that is, $\rho = \rho (r,t)$. This is a natural hypothesis, as it reflects the space isotropy. Accordingly, the energy current density $\vec{j}$ must be spherically symmetric too, that is, $\vec{j}=\vec{j}(r,t)\hat{r}$. Therefore, equation \eqref{noncont+rho} turns into

\begin{equation}\label{eq_div_j}
\frac{1}{r^2}\frac{\partial}{\partial r}\left( r^2j(r,t)\right)=k\rho(r,t)-\frac{\partial \rho(r,t)}{\partial t}.
\end{equation}
  
Integrating over the spatial variable $r$ we find
\begin{equation}\label{jota_integrado}
j(r,t)=\frac{1}{r^2}\left[A(t)+\int\left(k\rho(r,t)-\frac{\partial\rho(r,t)}{\partial t}\right)r^2dr\right],
\end{equation}
where $A(t)$ plays the role of constant of integration (with respect to the radial variable).
The corresponding differential equation for $f_T$ reads
\begin{equation}\label{new_diff_eq_f}
j(r,t)\,\frac{\partial f_T}{\partial r}+
\left(k\rho(r,t)-\frac{1}{2}\frac{\partial\rho(r,t)}{\partial t}\right)f_T=-8\pi k \rho(r,t).
\end{equation}

Equation \eqref{new_diff_eq_f} can be integrated by standard methods and its general solution is given by
\begin{equation}\label{solucao}
  f_T(r,t)=\frac{-8\pi k \displaystyle\int \frac{\rho(r,t)}{j(r,t)}\exp\left[\int\left(\frac{k\rho(r,t)-\frac{1}{2}\frac{\partial \rho(r,t)}{\partial t}}{j(r,t)}\right)dr\right]dr+B(t)}{\exp\left[\displaystyle\int\left(\frac{k\rho(r,t)-\frac{1}{2}\frac{\partial \rho(r,t)}{\partial t}}{j(r,t)}\right)dr\right]},
\end{equation}
where $B(t)$ is another ``constant of integration".

A special case of interest, motivated by the cosmological consideration of homogeneity of the Universe, comes from supposing that $\rho = \rho(t)$. Using this information in equation \eqref{jota_integrado} we find
\begin{equation}\label{result_current}
j(r,t)=\frac{1}{r^2}\left[A(t)+\frac{1}{3}\left(k\rho(t)-\frac{d\rho(t)}{dt}\right)r^3\right].
\end{equation}
Substituting equation \eqref{result_current} into \eqref{solucao} and performing the necessary integrations we get
\begin{equation}\label{special_case}
f_T(r,t) = B(t)\left(A(t)+\frac{\left(k\rho(t)-\frac{d\rho(t)}{dt}\right)}{3}r^3\right)^{\displaystyle\left(\frac{\frac{1}{2}\frac{d\rho}{dt}-k\rho(t)}{k\rho(t)-\frac{d\rho}{dt}}\right)}-\frac{8\pi k\rho(t)}{k\rho(t)-\frac{1}{2}\frac{d\rho(t)}{dt}}.
\end{equation}

It is interesting to analyze the behaviour of \eqref{special_case} when the system approaches the equilibrium, that is, $\frac{d\rho(t)}{dt}\rightarrow0$ and $\rho(t)\rightarrow\rho_{eq}=\text{cte}$. In this regime we find
\begin{equation}\label{equilibrium}
f_T(r,t)\rightarrow\frac{B(t)}{\displaystyle\left(A(t)+\frac{k\rho_{eq}r^3}{3}\right)}-8\pi.
\end{equation}
We note that the solution \eqref{equilibrium} is well-behaved at spatial infinity (for finite $t$), since $\lim_{r\rightarrow \infty}f_T(r,t)=-8\pi$. Therefore, at least for this asymptotic region, one can write 
\begin{equation}
f_T=\frac{df(T)}{dT}\approx -8\pi \Rightarrow f(T)\approx -8\pi T.
\end{equation}
(Notice that in a flat background the Ricci scalar is null, so that we can write $\frac{\partial f(R,T)}{\partial T}=\frac{df(T)}{dT}$.)

We highlight that this solution was obtained for a perfect fluid in a flat homogeneous and isotropic background, similarly to other studies on flat spaces. For example, in \cite{harko2011} it was shown the possibility of reconstruction of FLRW cosmology and its implications from $f(R,T)$. In \cite{shabani2013} the cosmological eras were studied.

%%%%%%%%%%%%%%%%%%%%%%%%%%%%%%%%%%%%%%%%%%%
\section{Conclusions}\label{sec:conc}
%%%%%%%%%%%%%%%%%%%%%%%%%%%%%%%%%%%%%%%%%%%

Nonconservation of energy in $f(R,T)$ gravity follows from the field equations and it is an intriguing feature of this theory. It occurs even in a flat spacetime background. It is interesting to compare this fact with the case of GR, where in asymptotically flat spacetimes the energy conservation law is well-established.

Within noncommutative theory we can also have energy nonconservation as a consequence of time discretization, which arises in the case of a compact extra dimension which does not commute with the time coordinate. In \cite{Balachandran2004,Balachandran2007} it was shown that the energy nonconservation should show up in scattering and decay processes. 

According to the model studied in the present work, the energy nonconservation in the 4-dimensional $f(R,T)$ theory can be understood as an energy flow from the ordinary (4D) spacetime into the compact extra dimension (and vice versa). In the 5-dimensional spacetime the total energy is conserved. The observer is not directly aware of the extra dimension, because of its smallness. On this sight, one can interpret the effect of a compact extra dimension as a breakdown of the conservation of the full energy-momentum tensor $T^{\rm full}_{\mu\nu}$ accompanied by the emergence of an effective (4D) nonconserved energy-momentum tensor $T^{\rm eff}_{\mu\nu}$.

The above mechanism is similar to the one discussed in \cite{Arkani-Hamed1998}, where there are $n\ge 2$ compact dimensions and energy is carried away towards the extra dimensions. We consider the low energy limit, when matter decouples from gravity, differently from \cite{Harko2015a}, where it was considered a matter-geometry coupling with energy-momentum exchange between them.

From an empirical perspective, we notice that the Large Hadron Collider (LHC) is currently looking for  phenomena beyond the Standard Model, in different energy scales (6-12 TeV). Certainly a violation of the conservation of energy-momentum in a particle process would be astonishing, forcing us to review the basic aspects of physics. If that happens, our model could be used to test the consequences of such a scenario for gravity theory. Specifically, one could try to fit the experimental data and measure $\kappa$, thus finding an estimate for the ratio $\theta/\mathcal{R}$. This would be valuable for other High Energy Physics models, which consider noncommutative physics with discrete time evolution.

Summarizing our results, we have obtained a modified continuity equation, showing that the energy conservation is not respected in $f(R,T)$ gravity in a four dimensional flat ($R=0$) space-time. We interpreted this feature as a consequence of the existence of a noncommutative compact extra dimension. This nonconservation of energy is caused by quantum transitions which do not respect the usual energy shell condition of ordinary quantum mechanics, and this is a consequence of the time discretization, a characteristic of the noncommutative model adopted. Energy is thus conserved only in the 5-dimensional space-time. In our model the energy flux is proportional to the energy density of the matter content, leading to a differential equation for $f_T(r,t)$. Assuming that the energy density is known, we solved the equation and found the general form of $f_T(r,t)$. Besides, we also considered the special case of a homogeneous energy density, which can be interesting for cosmological applications. We wish to explore this possibility in a forthcoming work.

\section*{Acknowledgements}
  RVL thanks CNPq (Conselho Nacional de Desenvolvimento Cient\'ifico e Tecnol\'ogico) process 141157/2015-1 and CAPES/PDSE/88881.134089/2016-01 for financial support.
  GAC thanks to CAPES (Coordena\c c\~ao de Aperfei\c coamento de Pessoal de N\'ivel Superior) for financial support process \# 88881.188302/2018-01.
  PHRSM would like to thank S\~ao Paulo Research Foundation (FAPESP), grant 2015/08476-0, for financial support. The authors would like to thank the referees for valuable help in optimizing the presentation of this work.

\bibliographystyle{unsrtnat}
\bibliography{library}

\end{document}